\documentclass[conference]{article}
\usepackage{cite}
\usepackage{amsmath,amssymb,amsfonts}
\usepackage{algorithm}
\usepackage{algpseudocode}
\usepackage{adjustbox}
\usepackage[graphicx]{realboxes}
\usepackage{textcomp}
\usepackage{xcolor}
\usepackage{rotating}
\usepackage{pdflscape}
\usepackage{threeparttable}
\def\BibTeX{{\rm B\kern-.05em{\sc i\kern-.025em b}\kern-.08em
    T\kern-.1667em\lower.7ex\hbox{E}\kern-.125emX}}
\usepackage{booktabs}

\begin{document}

\title{FGAN: Federated Generative Adversarial Networks for Anomaly Detection in Network Traffic}

\author{Sankha Das\vspace{2 mm}\\
BITS Pilani, Rajasthan, India.\\
f20190029@pilani.bits-pilani.ac.in}
    

\maketitle

\begin{abstract}
    Over the last two decades, a lot of work has been done in improving  network security, particularly in intrusion detection systems (IDS) and anomaly detection. Machine learning solutions have also been employed in IDSs to detect known and plausible attacks in incoming traffic. Parameters such as packet contents, sender IP and sender port, connection duration, etc. have been previously used to train these machine learning models to learn to differentiate genuine traffic from malicious ones. Generative Adversarial Networks (GANs) have been significantly successful in detecting such anomalies, mostly attributed to the adversarial training of the generator and discriminator in an attempt to bypass each other and in turn increase their own power and accuracy. However, in large networks having a wide variety of traffic at possibly different regions of the network and susceptible to a large number of potential attacks, training these GANs for a particular kind of anomaly may make it oblivious to other anomalies and attacks. In addition, the dataset required to train these models has to be made centrally available and publicly accessible, posing the obvious question of privacy of the communications of the respective participants of the network. The solution proposed in this work aims at tackling the above two issues by using GANs in a federated architecture in networks of such scale and capacity. In such a setting, different users of the network will be able to train and customize a centrally available adversarial model according to their own frequently faced conditions. Simultaneously, the member users of the network will also able to gain from the experiences of the other users in the network. The training is performed in a coordinated and localized setting, which will also eliminate the need to collect data from users into a central storage and ensure privacy. On a whole, the proposed work introduces an architecture and mechanism for a well-coordinated intelligent intrusion detection system, addressing concerns of large-scale shared model updates, date privacy and model tampering.\vspace{1 mm}
\end{abstract}

\section{Introduction}
Computer and mobile networks have evolved enormously over the last three decades, in both size and capabilities. As more number of nodes are able to join and communicate across a network, the internal and external traffic in these networks has also grown substantially over the years. The traffic in large networks varies in both capacity and a number of physical features such as bandwidth, frequency, connection duration, communication protocol, etc. Parameters which change from connection to connection include sender's IP address, sender's port, receiver's IP address, receiver's port, packet headers, packet contents, protocol, etc. Packet detection is achieved by analysing these payload and protocol data. Such varied network traffic is commonly observed in medium to large-scale networks such as offices and metropolitan area networks. It is therefore clear that traffic in such networks provide a large multitude of parameters to study and analyse the traffic. 

However, the large capacity of these networks also acts as a vulnerability. The sheer amount of different communicating parties that the nodes of the network interact itself is an indicator that a substantial part of this population might be attackers trying to send malicious traffic to the network. The analysis of such heavy-volume traffic for effective anomaly detection brings with itself some inherent issues of dynamic nature of the network, adaptability of the detection systems to modern attacks, the diversity of the attacks and accuracy of the detection \cite{shone2018deep}. In such cases, the use of machine learning (ML) methods such as deep belief networks, support vector machines, decision trees and other statistical tools for intrusion detection have proved to be helpful \cite{dong2016comparison}. Limitations still exist with such methods, including concerns about training time and data overhead, accuracy of detection and false positives. However, the  adaptability and versatility of such models for different kinds of network traffic make them a very popular choice for building intrusion detection systems. The contemporary scenario of using ML-based techniques in network security are more oriented towards handling a large and more varied range of attacks. As the form and functions of computer networks have evolved over time, so have the mechanics of the attacks that are posed on these networks. Nowadays, common attacks target vulnerabilities of the networks such as the network architecture, routing algorithm, communication frequencies and ports, etc. However, different kinds of attacks may have completely unrelated feature vectors, which becomes a weakness of traditional ML-based techniques \cite{elsayed2020detecting}.

This paper proposes the architecture and functioning of an intrusion detection system (IDS) to detect malicious traffic entering a network. An intrusion detection system is entrusted with the task of monitoring activity and data exchanges taking place in a device or a network or devices. The system should flag and alert against any malicious activity that might be detected so that the device or network may take appropriate corrective actions. Intrusions may be detected by comparing activity with a database of known attacks (misuse detection) or by detecting any deviation from the normal activity occuring otherwise on the system or network (anomaly detection) \cite{4667434}.  Intrusion detection systems are broadly classified into host intrusion detection systems (HIDS) and network intrusion detection systems (NIDS).  HIDSs monitor the activity only for a single system, such as file exchanges or tasks taking place for a single device. NIDSs on the other hand monitor and track activity for an entire network of connected devices. An NIDS is capable of monitoring traffic being exchanged by and within the network. The solution proposed in this work falls under the category of NIDSs, where it can handle activity monitoring for small to medium and large-scaled networks. The proposed IDS will be using a Generative Adversarial Network to detect deviations of the network traffic from normal by classifying them as anomalous. This detection will also be used to further train the existing model to make it adaptable for detecting a varied range of malicious traffic. Models trained at one location of a large network will then be made available for use across the entire network through a central aggregated model which will be constituted by taking into account the updates made to the model at each node of the network. This update and aggregation process will be coordinated by servers designated at each level of the network, exchanging data with the rest of the network through encrypted communications. The main idea behind developing such a distributed intrusion detection system is to enable using the techniques used to detect malicious activity at one location of the network in other parts as well. Additionally, it is also aimed that the IDS is able to detect new attacks by building on the experience gained in the past and other parts of the network.

The remainder of this paper is arranged as follows. Section II discusses the necessary primitives required to understand the idea proposed in this work, namely on the topics related to network security, anomaly detection in networks, GANs and Federated Machine Learning. Section III discusses the work done previously in the domain of intrusion detection in networks. Section IV presents the Federated Learning architecture powered by GAN models. The proposed architecture and its functions can be used to detect anomalies in network traffic to determine whether a given packet is legitimate or malicious. Section V summarizes the work and discusses a few scopes for future research and extension of the proposed work.

\section{Background}
This section discusses the necessary primitives pertaining to intrusion detection systems, anomaly detection, machine learning and their relevance to network security. The concepts emphasised here will be helpful in understanding the content of the proceeding sections on related work in ML-based anomaly detection and the proposed scheme of federated GANs for anomaly detection. 

\subsection{Network Security and Intrusion Detection Systems}
Communications continue to take place in and across networks, often with new and unknown nodes. Securing the network against malicious communications and preventing the compromise of data pertaining to the users in the network are some of the objectives of ensuring network security. The knowledge of the underlying network architecture and the commonly observed parameters in its communications are often targeted by attackers to direct malicious traffic into the network. The job of network intrusion detection systems (NIDS) is to detect such malicious and anomalous traffic and warn the network against them. Most IDSs do so by monitoring the network traffic through resources such as system commands and system logs \cite{intrusionSurvey}. While anomaly detection is mostly associated with looking for unusual behaviour in network traffic suggesting possible novel attacks, signature detection measures check for changes in network traffic patterns which indicate previously known attacks \cite{mahoney2003learning}. Most anomaly detection schemes rely upon behavioral changes in network traffic from the regular traffic patterns observed in the network. It is expected that an ideal anomaly detector should be able to distinguish malicious traffic from legitimate traffic with high accuracy across different environments and network architectures.

A vital step in building anomaly detection systems is to decide upon the features or parameters to analyse in order to detect an intrusion. The authors in \cite{guyon2003introduction} emphasise on proper variable-selection to achieve better predictability of the proposed predictors and also improving their efficiency and speed. Speed is of importance in anomaly detection so as to prevent any loss to the network due to the intrusion. At the same time, knowledge of underlying data that is analysed by the predictors adapt to future attacks which may have subtle changes in the underlying feature vectors \cite{iglesias2015analysis}. Different kinds of attacks such as network probing, denial-of-service (DoS), Root to Local (R2L) and User to Root (U2R) are common in most networks and each of them entail a different set of attributes that should be targeted by IDSs to detect their presence in incoming traffic. Additionally, proper selection of attributes for developing anomaly detectors is necessary from the perspective of effective machine learning and statistical observations. As the authors cite in the latter reference, the consideration of excess and redundant variables in the feature vector result in performance degradation of a number of anomaly classification techniques such as Support Vector Machines (SVM) and k-Nearest Neighbours (kNN) classifiers. Correlated variables impose noise upon techniques such as Naïve Bayes and LAR models. The infamous curse of dimensionality associated with large-dimensional feature vectors is yet another reason for the importance of feature reduction in anomaly detection systems \cite{zimek2012survey}.

\subsection{Generative Adversarial Networks}
Generative Adversarial Networks (GAN) were introduced in the work of Ian Goodfellow et at. \cite{goodfellow2014generative} in 2014 and since then have found applications in multiple domains such as image processing, speech synthesis, anomaly detection, etc. GANs are composed of a generator model $G$ and a discriminator model $D$ which are deep neural networks that are trained in an adversarial setting. The generator, as the name suggests, generates instances of fake data which try to mimic the underlying distribution of the dataset. The discriminator is tasked with the job to distinguish these fake data instances from the genuine instances when tested on a random mixture of both. The generator hence tries to fool the discriminator by attempting to generate data which cannot be distinguished from the real data. If a stage is reached when the discriminator can predict the genuineness of the input data instance with only $1/2$ probability, then the generator is said to have achieved maximum performance as now the fake and genuine data are indistinguishable. A common analogy often cited to distinguish between the two models is that of a con and cop - the generator is analogous to the con, trying to befool and escape the discriminator (the cops) by producing counterfeit notes; the cops try to distinguish the fake notes from the genuine ones and hence try to identify the con. The process of training a GAN entails the objective of finding a set of optimal parameters for the generator and discriminator networks: the generator using the parameters of its model to mimic the real data distribution and the discriminator classifying the actual data samples from the synthesized ones. Like all other machine learning algorithms, this training also calculates a loss to evaluate the accuracy of the generator and discriminator. The loss is defined as the generator-discriminator dependent value function $V(G, D)$, given by

\begin{center}
    \( V(G, D) = E_{p_{data}(x)} log(D(x)) + E_{p_g(x)} log(1 - D(x))\)
\end{center} 

$V(G, D)$ is the cross-entropy loss function between the generator and discriminator models, with the generator trying to minimize the value of $V(G, D)$ and the discriminator trying to maximise it. Here $E_{p_{data}(x)} log(D(x))$ represents the expectation of the event that the a certain data sample $x$ was classified by the discriminator as coming from the actual data distribution. $E_{p_g(x)} log(1 - D(x))$ represents the expectation of the event that the a certain data sample $x$ was not classified as being generated by the generator. Therefore, the objective of training a GAN is to find the optimal set of parameters which satisfies $\min_{G} \max_{D} V(G, D)$. The training process involves generating the fake samples by the generator and combining them with the real samples. The discriminator classifies these two sets from each other, the classification process updating the parameters of both generator and discriminator to satisfy the min-max condition of $V(G, D)$.

\begin{figure*}[htp]
    \centering
    \includegraphics[width=12cm]{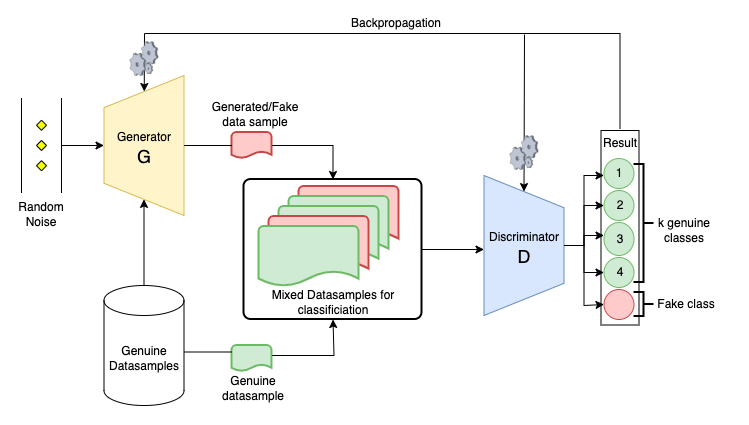}
    \caption{A Generative Adversarial Network Architecture}
    \label{fig:training clustering}
\end{figure*}

\subsection{Federated Machine Learning}
The federated learning paradigm was proposed in a number of papers published by researchers at Google \cite{DBLP:journals/corr/KonecnyMRR16} \cite{mcmahan2017communication} as a solution for distributed and localized model training on mobile devices. The training data and training process is localized in each mobile device, and only the trained model's updates are sent to a centralized server for aggregation. Therefore, training data and all kinds of information, public or private, is concealed from the external world and only the trained model is exposed. This approach has significantly addressed the controversial views over the exposure of sensitive data through centralized training datasets across different domains. Especially in situations such as medical and financial records which might be too confidential or personal to compromise, the federated learning paradigm uses the data for training a shared model without the need to expose the underlying dataset. The objective of a federated learning process is to train a centrally coordinated and shared model by aggregating updates learned locally on different devices. Intuitively, this gives the idea of combining the model parameters collected from different devices by means of a weighted average in order to produce the final optimal model. The final model can then be shared again with the participating devices for another round of model updates, with the devices now training the newly updated model. As discussed above, federated learning addresses security and data privacy issues through localized training data and encrypted model sharing. In addition, it also solves the issue of handling large amounts of training data at one centralized device. Model training takes place independently across a number of devices, often giving rise to the possibility of training using varied algorithms and datasets.

\section{Related Work}
Traditional ML-based methods are not suitable for handling dynamically evolving attacks and are not suited for non-linear datasets. \cite{elsayed2020detecting}. Many previous works have also proposed the use of artificially generated attacks and their classification for developing robust intrusion detection systems. It has been studied before that the parameters used to predict a potential anomaly can differ from attack to attack, and are hence dynamic in nature \cite{mishra2018detailed}. The right choice of features that affect or characterize an attack needs to be identified in order to produce artificially generated attacks for improvising network security. Authors in \cite{charlier2019syngan} have used a GP-WGAN model for generating attacks that have achieved errors levels as low as 0.10 root mean squared, including generation of DDoS attacks. A GAN-based model IDSGAN has been proposed in \cite{lin2018idsgan} which was tested on the NSL-KDD dataset and classifies time-based, intrinsic, content-based and protocol oriented attacks such as probing, DoS, U2R and R2L attacks. A similar model can be used in a federated learning setup to suit the solution being proposed in this work, with the learning and updation of parameters in the generator and discriminator taking place in a localized and distributed manner. The work in \cite{zhao2020intelligent} discusses the use of LSTMs in a federated learning setup for distributed learning and centralized model updates for intelligent intrusion detection systems. Therefore, a significant amount of work has been done in the past to explore the possibilities of adversarial attacks and their detection, in both centralized and distributed settings. The idea proposed in this work seeks to build upon these works and propose a federated learning architecture and functionality for detecting diverse anomalies in medium to large-sized networks in a coordinated manner.

\section{Federated GAN}
This section details about the form and function of the proposed solution of Federated GANs. From a bird's eye view, this solution has a number of GANs operating at different regions of the network to detect anomalous traffic; in addition, the parameters learned in the process of individual training of each GAN are shared with the rest of the GANs. As a result, each region of the network is able to access a shared model which can be used for anomaly detection in network traffic. One of the main objectives behind proposing a federated learning setting for GANs in anomaly detection is to include the characteristics of model sharing and dataset privacy. The subsections below discuss the architecture, training process and coordinated working of each of the GANs in the network to achieve the target of the network.

\subsection{Architecture}
For the purpose of explaining the architecture of the model, let us consider a large internal network inside an organization. Let this network be composed of smaller networks of nodes which are exposed to similar nature of traffic for the majority of their operation. We will call these smaller networks as training clusters. For simplicity, let us consider these smaller networks to be served by a single switch or proxy server. All traffic that passes the nodes in these smaller networks, pass through the corresponding proxy server. The larger network will be composed as a network of all such proxy servers. A central server connects all the proxy servers and performs tasks of updating and coordinating the shared model within the network. Therefore, the overall network is represented as a federation of devices updating and using a common shared model. This shared model will be used to detect anomalous network traffic entering the network. Different regions of the network may be using different physical media for communication. This may range from co-axial cables and fibre optic cables to wireless connections.

\begin{figure*}[htp]
    \centering
    \includegraphics[width=12cm]{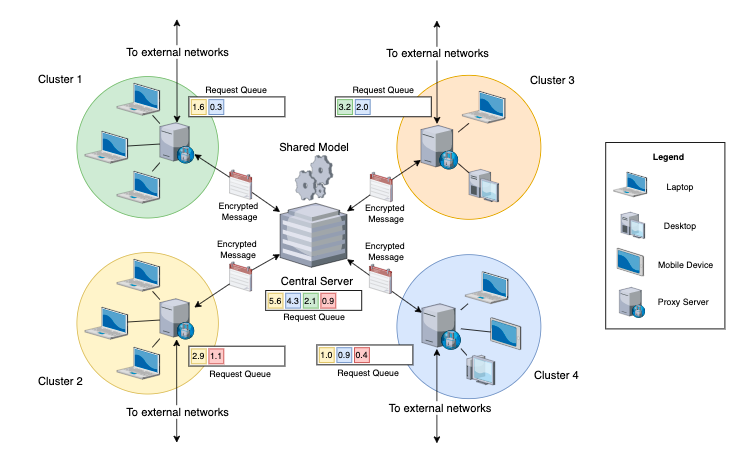}
    \caption{Proposed FGAN Architecture}
    \label{fig:training clustering}
\end{figure*}

Each of the proxy server maintains a list of the nodes it serves in its internal network. While training of the model takes place in a localized manner in each node, these nodes push the trained model's updates to their respective proxy server. The proxy server in turn pushes aggregated model updates to the central server. The central server, after incorporating updates from all proxy servers in a given period of time, makes the shared model available to the entire network. A copy of the updated shared model will be downloaded by each proxy server again, which in turn makes it available to the nodes in its corresponding training cluster. The detailed process of how the shared model is trained and coordinated is detailed in the following two sub-sections. 

\subsection{Training the Model}
The model that will be shared and trained across the network is a generative adversarial network. The training process will require jointly training the generator and the discriminator until both attain a sufficiently high degree of accuracy in outsmarting the other (ideally speaking, the probability of the discriminator detecting a fake sample generated by the generator should be exactly 0.5). In our use case, it suffices to train the model locally at each node using a method like stochastic gradient descent (SGD) and backpropagation. The input vector to the neural network will comprise of parameters important for ascertaining the nature of the traffic as genuine or malicious. The training dataset needs to be created by the respective node using the incoming traffic. All incoming traffic data is logged locally in the node, and the nature of the traffic is ascertained by the node by checking for any harmful effects caused by it. A traffic vector is created using the parameters of the incoming traffic, and a tag is attached to it indicating the vector as either genuine or malicious. The generator would then generate samples of such traffic vectors resembling the distribution of the input dataset by introducing a noise vector. During the training process, the generator keeps generating fake samples of data and tries to minimize the generator loss $L(G)$, while the discriminator tries to correctly distinguish the fake samples from the real ones, thereby reducing the discriminator loss $L(D)$. Once the training phase is over, the generator is separated from the model and the trained model is tested over incoming traffic. The accuracy of the model is now defined by the probability of the discriminator successfully distinguishing anomalous traffic from genuine traffic. Once the model has been trained, the updates for the model are pushed to the proxy server of the training cluster at a time convenient to the node (generally at a time when the node is idle or not performing traffic heavy or resource intensive tasks).

The update requests received by the proxy server from the nodes now have to be used to generate an aggregated model. This is done using an average aggregation algorithm such as FedAvg \cite{mcmahan2017communication}, where the final updated parameters are generated by taking a weighted average of the gradient updates submitted by each individual node. FedAvg defines the predicted loss from a single node for given model parameters $w$ and data set $P_k$ as 

\begin{equation}
    F_k(w) = \frac{1}{n_k} \sum_{i \in P_k}{f_i(w)}
\end{equation}

where $n_k = |P_k|$ and $f_i(w)$ is the loss predicted from the $i^{th}$ data sample. The aggregated loss derived by the server for updates submitted by $K$ nodes is given by

\begin{equation}
    f(w) = \sum_{i = 1}^{K} {\frac{n_k}{n} F_k(w) }
\end{equation}

In the proposed solution, an additional impact parameter $h$ is introduced for each request submitted by the corresponding node. This impact parameter is used to associate a weight in addition to the weights introduced by the FedAvg algorithm. The reasons of introducing the impact parameter for taking a weighted average will be explained in the next subsection. Therefore, the modified equation for calculating the aggregated loss is given by

\begin{equation}
    f_{FGAN}(w) = \sum_{i = 1}^{K} {\frac{n_k}{n} *  h_i * F_k(w) }
\end{equation}

\subsection{Coordination}
The application of the proposed solution can be realized in networks of varying sizes. In the case of small-sized local networks, managing the training process of the shared model and its subsequent distribution is a relatively easy task. This can be accounted for by the low variance in the nature of the network traffic and lesser number of model updates required. In such a scenario, the network administrator may also chose to the skip the inclusion of the proxy servers in the architecture and coordinate all model updates and distribution directly between the central server and nodes at the network edge. The nodes train the shared model locally, and push the trained model's updates to the central server. The central server updates the shared model on the fly as newer updated models are received from the nodes, and makes it available for the network. In such a case, the central server may not require any kind of scheduling mechanism for making updates to the shared model. However, as we begin to consider medium to large-sized networks such as the one mentioned in subsection IV.A, things start to become more complicated. Firstly, in such networks a large number of training clusters may be connected with each other. A large number of nodes may be operating within each of these training clusters, served by their respective proxy server. Each of these nodes will be locally training a copy of the shared model and pushing their updates to the proxy server. Therefore, the first level of coordination needs to be implemented at the proxy servers. Secondly, the large size of the network also makes it exposed to a more varied nature of traffic. Each of the training clusters may be exposed to different sources and thus, to a possibly wide nature of traffic. The third difficulty arises at the level of the central server, which needs to update the shared model based on the training updates received from a possibly large number of training clusters. Since continuously updating the shared model in response to a large number of updates can be a huge performance degradation, the central server also needs to employ some scheduling mechanism to efficiently update the shared model. 

A priority queue is used for scheduling the nodes in each training cluster for pushing the model's updated gradients to the respective proxy server, and to let proxy servers push their updates to the central server. As the requests for submitting updates arrive at a server (either the proxy server or central server), they are ordered according to the decreasing priority of the requests in the queue. Let us first examine the scheduling mechanism at the proxy server. Each request will have a priority value $p$, which will be used for ordering the requests in the priority queue. Let there be $N$ nodes in a given training cluster apart from the proxy server. We define $C$ as the fraction of nodes in the cluster whose update requests are considered during a model update round at the proxy server. Therefore, the top $\lfloor C*N \rfloor$ requests in the queue based on priority value $p$ are considered in a given model update round. The priority value $p$ is calculated as follows:

\begin{equation}
    p = \frac{A}{N * (\frac{T - T_s}{T - T_o})}
\end{equation}

Here, A is defined as the attack index of the node which indicates the count of malicious responses that the node has received since it joined the cluster. The higher the attack index of the node, the more priority its model updates should be given while generating the aggregated model. The term $(\frac{T - T_s}{T - T_o})$ indicates the the maturity index of the node, defined as the ratio of $ T - T_s $, the time elapsed since the node joined the cluster and $ T - T_o $, the time elapsed since the creation of the cluster. The maturity index lowers the priority of newly added nodes in the cluster in an attempt to reduce the clutter from high rate of node addition and request submission. This also mitigates the risk of malicious nodes trying to join the cluster frequently and trying to submit spurious model update requests to the proxy server. $N$ as defined above, indicates the number of nodes in the network, and produces a uniform distribution effect of priority values among all nodes in the cluster. Static values such as $T_o$, $T_s$ and $N$ are stored in the proxy server. The value of $A$ is maintained by each node and submitted to the proxy server along with the model update request. The priority value of an update request is calculated by the proxy server at the time when the request is added to the priority queue after being submitted by a node. Here, there is a possibility that a node in order to increase the priority value of its request reports a high value of $A$ to the proxy server while submitting the request. To discourage such activity, a node is blacklisted from the cluster for a stipulated amount of time if it constantly reports a high value of $A$ for 3 consecutive update requests. This threshold value may be set by each proxy server according to the cluster size and the cumulative traffic volume that the cluster receives. If such a case arises, the node is blacklisted from the cluster for a stipulated suspension time $T_{sus}$ and its joining time $T_s$ is reset to the time when its suspension period ends. This penalization strategy also produces a precautionary effect on the cluster in cases of high rate of actual attacks on a node in the cluster. By blacklisting the node from the cluster, the traffic exchanged between the node and the external network is stopped and the probability of attacks on the internal network is reduced for the suspension period. This is analogous to the precautionary measure of closing down a road or bridge on which a high number of accidents have been reported recently. 

The nodes in a cluster may submit a request for model updates whenever they wish. However, they can submit at most one request until the next aggregated model has been generated by the proxy server. Therefore, at most one request can be submitted by a node for the next aggregated model update. The commands for training the model and pushing model updates are not initiated by the proxy server as in conventional federated learning and nodes are given complete autonomy for training the model and pushing their respective updates to the server. The proxy server chooses a convenient time (typically when it is mostly idle and network traffic is low) for performing the model aggregation based on the requests submitted by the nodes in the cluster. The proxy server takes the first $\lfloor C*N \rfloor$ requests in the priority queue, and discards the rest of the requests, making the queue empty. It then performs the model aggregation using the algorithm described in subsection IV.B. Here we pass the impact parameter discussed there for performing weighted averaging of the model updates to produce the aggregated model. The impact parameter is nothing but a vector of the priority values of the respective requests being considered for the updating the model. Therefore, the coordinated model update takes place following the below algorithm.

\begin{algorithm}
  \caption{Coordinated Model Update}\label{coord-model}
  \begin{algorithmic}[1]
    \Procedure{model-update}{$model, queue, C, N$}
      \State $m \gets \lfloor C*N \rfloor$    
      \State $updates\gets [1..m]$
      \State $impacts\gets [1..m]$
      \For {i = 1 \textbf{to} m}
        \State $req \gets queue.EXTRACT-TOP()$
        \State $updates[i] \gets req.GET-UPDATES()$
        \State $impacts[i] \gets req.PRIORITY()$
      \EndFor
      \State $queue.EMPTY-QUEUE()$
      \State $model \gets AVG-AGGR(model, updates, impacts)$
      \State \textbf{return} $model$
    \EndProcedure
  \end{algorithmic}
\end{algorithm}

Here the aggragation procedure $AVG-AGGR()$ is similar to the FedAvg algorithm as explained in subsection IV.B. Once the aggregated model is generated by the proxy server, two tasks are executed: 1) pushing a copy of the updated model parameters to the central server, and 2) distributing the updated model to all the nodes in the training cluster. On receiving the updated model, the nodes in the cluster start using this model for detecting anomalous traffic and training it.

The coordination process at the central server takes in a similar fashion. The only differences introduced are that the central server receives updates from proxy servers of the training clusters instead of the individual nodes. The proxy servers push the aggregated models they have produced locally to the central server in the form of requests similar to the requests submitted by the individual nodes. Every request submitted by a proxy server has a priority value $P$ given by:

\begin{equation}
    P = \frac{A_C}{N_C * (\frac{T - T_s}{T - T_o})}
\end{equation}

where $A_C$ is the attack index of the training cluster, given the sum of the attack indices of all nodes in the cluster. $N_C$ indicates the number of training clusters in the network and $T_s$ and $T_o$ represent the time of creation of the cluster and the time of creation of the network respectively. The logical meaning of the terms remain same as the case of the proxy server level coordination. The central server generates the aggregated model from the update requests submitted by the different training clusters using the same weighted average aggregation method as in the case of updating the model at the proxy servers. In the case of the central server, the impact parameters are given by the priority values of the request submitted by the respective training cluster. Once the aggregated model has been generated by the central server, it is distributed to all training clusters through the proxy servers. The nodes in each cluster now store and train this newly generated model and discard all copies of any previous models. This new model incorporates the changes introduced by the training process on highly varied traffic across different clusters in the network. Therefore, this enables all nodes in the network to detect a range of diverse anomalous traffic by sharing and exchanging information in form of the shared model. 

\subsection{Security and Privacy Concerns}
As highlighted in \cite{DBLP:journals/corr/KonecnyMRR16}\cite{mcmahan2017communication}, the federated learning paradigm introduces solutions to many security and data privacy issues associated with conventional machine learning techniques such as centralized data storage, compromise of sensitive data embedded in training data, etc. Use of federated learning for coordinating model updates in the proposed solution addresses the same issues, with no requirement of tracking the communications of each individual node in a training cluster. As a result, privacy of data exchange and sensitive information that might be enclosed in the traffic payload is not compromised. In conventional practices, the model updates communicated between the server and nodes is encrypted to prevent unauthorized access and changes to the updates. In the baseline situation of the server and all nodes being legitimate, the encrypted data can be decrypted and used to update the model without expecting any malicious tampering with the model updates by the server. However, let us consider the case of the server itself being malicious (which may be instigated by any internal or external agents). In this case, the server may change the gradient updates submitted by the nodes in the training cluster, resulting in an inaccurate final aggregated model. This inaccuracy has a number of consequences: (1) this model might fail to correctly classify malicious traffic from legitimate traffic in certain cases; (2) the wrong parameters of the model will be used in the next rounds of training by the nodes in the training cluster, thereby propagating the inaccuracy in the further rounds; (3) in the case that the server is selectively ignoring updates submitted by certain nodes, the model may become highly biased towards classifying only a certain kind of traffic, leading to inefficiency and further inaccuracy. To solve this issue, it is proposed to use homomorphic encryption techniques \cite{ogburn2013homomorphic}\cite{gentry2011implementing}\cite{fontaine2007survey} for encrypting and operating on the model updates exchanged between the server and nodes. Using homomorphic encryption, the server can produce the aggregated model by operating over the encrypted gradients submitted by the nodes, without the need to decrypt them. The final aggregated model produced will have its parameters encrypted as well, which will be used in further rounds of training by the nodes in the training clusters. As a result, the risk of tampering with the model updates or the final aggregated model parameters will be reduced. 

\section{Conclusions}
The solution of Federated GANs for detecting intrusions in medium to large-scaled networks as introduced in this work will help to build a more secure and well-coordinated mechanism for an intelligent IDS. The impact parameter and penalties associated with the model updates submitted by each node in a training cluster help in adding priority to each request, thereby increasing fairness and legitimacy of update requests. A distributed setting of coordinating the model updates through training clusters, proxy servers and the central server helps to reduce the load on the central server as each training cluster's updates are taken care of by the corresponding proxy server. As a scope of future research, the architecture may be further improved by introducing parallel central servers and introduction of peer-to-peer communications between training clusters and proxy servers. Many security and data privacy concerns are addressed in this work, which may be further improvised in the future. A number of tasks may be explored as future research topics such as assessing the integrity of the algorithms being used in the training or model aggregation process, and securing the process of adding new nodes to a training cluster.

\end{document}